\documentclass[aps,prd
,preprint,tightenlines,nofootinbib,showpacs]{revtex4}
\usepackage{amssymb,latexsym}
\usepackage{amsmath,amsbsy,bbm}
\usepackage{epsfig,bm}
\usepackage{graphicx,comment}
\DeclareMathOperator{\tr}{tr}

\begin{document}
\def\a{{\alpha}}
\def\b{{\beta}}
\def\d{{\delta}}
\def\D{{\Delta}}
\def\e{{\varepsilon}}
\def\g{{\gamma}}
\def\G{{\Gamma}}
\def\k{{\kappa}}
\def\l{{\lambda}}
\def\L{{\Lambda}}
\def\m{{\mu}}
\def\n{{\nu}}
\def\o{{\omega}}
\def\O{{\Omega}}
\def\S{{\Sigma}}
\def\s{{\sigma}}
\def\th{{\theta}}

\def\ol#1{{\overline{#1}}}

\def\Dslash{D\hskip-0.65em /}
\def\Dtslash{\tilde{D} \hskip-0.65em /}
\def\sumint{\sum \hskip-1.35em \int_{ \hskip-0.25em \underset{\bm{k}}{\phantom{a}} } \, \, \,}

\def\CPT{{$\chi$PT}}
\def\QCPT{{Q$\chi$PT}}
\def\PQCPT{{PQ$\chi$PT}}
\def\tr{\text{tr}}
\def\str{\text{str}}
\def\diag{\text{diag}}
\def\order{{\mathcal O}}

\def\meff{{m^2_{\text{eff}}}}

\def\Meff{{M_{\text{eff}}}}
\def\cF{{\mathcal F}}
\def\cS{{\mathcal S}}
\def\cC{{\mathcal C}}
\def\cE{{\mathcal E}}
\def\cB{{\mathcal B}}
\def\cT{{\mathcal T}}
\def\cQ{{\mathcal Q}}
\def\cL{{\mathcal L}}
\def\cO{{\mathcal O}}
\def\cA{{\mathcal A}}
\def\cR{{\mathcal R}}
\def\cH{{\mathcal H}}
\def\cW{{\mathcal W}}
\def\cM{{\mathcal M}}
\def\cD{{\mathcal D}}
\def\cN{{\mathcal N}}
\def\cP{{\mathcal P}}
\def\cK{{\mathcal K}}
\def\Qt{{\tilde{Q}}}
\def\Dt{{\tilde{D}}}
\def\St{{\tilde{\Sigma}}}
\def\cBt{{\tilde{\mathcal{B}}}}
\def\cDt{{\tilde{\mathcal{D}}}}
\def\cTt{{\tilde{\mathcal{T}}}}
\def\cMt{{\tilde{\mathcal{M}}}}
\def\At{{\tilde{A}}}
\def\cNt{{\tilde{\mathcal{N}}}}
\def\cOt{{\tilde{\mathcal{O}}}}
\def\cPt{{\tilde{\mathcal{P}}}}
\def\cI{{\mathcal{I}}}
\def\cJ{{\mathcal{J}}}

\def\eqref#1{{(\ref{#1})}}
\preprint{UMD-40762-428}

\title{Volume Effects for Pion Two-Point Functions in Constant Electric and Magnetic Fields}

\author{Brian C.~Tiburzi}
\email[]{bctiburz@umd.edu}
\affiliation{Maryland Center for Fundamental Physics, Department of Physics, University of Maryland, College Park,  MD 20742-4111, USA}

\date{\today}

\pacs{12.38.Gc, 12.39.Fe}

\begin{abstract}
We compute finite volume effects 
relevant for lattice QCD simulations using background fields. 
Focusing on constant electric and magnetic fields on a periodic lattice, 
we determine volume corrections to pion two-point functions using chiral perturbation theory. 
Among such corrections are the finite volume shifts to the electric
and magnetic polarizabilities, which are numerically shown to be non-negligible.
We additionally find that all terms in the single-particle effective action
can be  renormalized by infrared effects. 
This includes Born couplings to the pion current and total charge-squared, which 
can be renormalized due to the nature of gauge invariance on a torus.
\end{abstract}

\maketitle

\section{Overview}

Lattice gauge theory simulations are making dramatic progress towards addressing quantitatively the non-perturbative dynamics of quarks and gluons%
~\cite{DeGrand:2006aa}. 
Despite its successes, 
lattice QCD suffers from systematic errors due to a number of sources. 
Understanding and accounting for these systematic errors is necessary for accurate determination of hadronic observables. 
In particular, 
as the lattice pion masses approach the physical pion mass
on a fixed-size lattice,
the effect of finite volume will become pronounced. 
This is increasingly important to note as lattice calculations are in 
reach of the physical point~\cite{Aoki:2008sm}.
Certain observables are more susceptible to volume corrections; 
thus, 
a practical application of field theories in finite volume
is the study of finite-size scaling of observables calculated using lattice QCD simulations.

The long-range physics of low-energy QCD is a consequence of the lightest degrees of freedom, 
the pseudo-scalar pions. 
In finite volume, 
hadronic observables are altered due to modified pion dynamics arising from boundary conditions imposed on the underlying lattice action. 
These finite-size effects can be treated systematically using chiral perturbation theory%
~\cite{Gasser:1987ah,Leutwyler:1987ak,Gasser:1987zq}. 
In this theory, 
the light pions emerge as a consequence of spontaneous chiral symmetry breaking,
and the smallness of the up and down quark masses compared to the QCD scale.
For QCD in the presence of external sources, low-energy hadronic properties are encoded in multipole moments, 
radii, and polarizabilities. 
These quantities are accessible with lattice QCD correlation functions using two different techniques:
the source insertion method and the classical external field method. 
The source insertion method does not directly probe such observables:
lattice data additionally require momentum extrapolation. 
Furthermore, at finite volume the multipole decomposition into momentum-transfer dependent 
form factors is no longer valid.
Finite volume studies of hadronic current matrix elements at non-vanishing momentum transfer and their generalizations 
have been studied using chiral perturbation theory~\cite{Chen:2006gg,Bunton:2006va,Hu:2007ts,Tiburzi:2007ep}.

In this work, 
we address finite volume effects relevant for QCD simulations in  background fields, see e.g.%
~\cite{Fucito:1982ff,Martinelli:1982cb,Bernard:1982yu,Fiebig:1988en,Aoki:1990ix,Christensen:2004ca,%
Lee:2005ds,Lee:2005dq,Shintani:2006xr,Engelhardt:2007ub,Detmold:2008xk,Aubin:2008hz}. 
Focusing on constant electric and magnetic fields, 
we use chiral perturbation theory to calculate pion two-point functions in the presence of classical external fields. 
These results can be utilized to analyze lattice correlation functions calculated in background field simulations.
Our computations demonstrate that the nature of gauge invariance on a torus leads to finite volume artifacts. 
A striking manifestation of this is the infrared renormalization of 
pion currents and charge-squared couplings. 
Identical effects have been found from momentum space calculations of current matrix elements~\cite{Detmold:2006vu,Hu:2007eb}.
Consequently two-point functions for both charged and neutral particles are modified at finite volume; 
for example, 
neutral particles propagating in external fields no longer have correlation functions with a simple exponential falloff at long times. 
On general grounds, 
one expects the finite volume renormalization of all terms in the effective action for a particle in a background field.

Our presentation is organized as follows. 
First in Section~\ref{s:Electric}, 
we consider the case of pions propagating in a background electric field. 
Here the field is treated in Euclidean space.
Corrections to the infinite volume single-particle effective
action are determined using spatially periodic boundary conditions. 
In Section~\ref{s:Magnetic}, 
we consider the case of pions propagating in a background magnetic field. 
The effects of finite volume are determined on 
the single-particle effective action for this case as well. 
For both electric and magnetic cases, we provide expressions
for the volume effects under the lattice approximation of 
omitting disconnected quark contractions. 
Finally in Section~\ref{s:summy}, 
we summarize our findings,
make important clarifications,
and discuss further applications of our work.

\section{Mesons in Finite Volume: Electric Case}
\label{s:Electric}

Pions propagating in a background electric
field at finite volume can be handled using Schwinger's
proper time method~\cite{Schwinger:1951nm}. 
We choose to implement the field with a time-dependent gauge potential
\begin{equation} \label{eq:AE}
A_\mu(x) = (0,0, - \cE x_4, 0)
,\end{equation}
which corresponds to a Euclidean electric field, 
$\cE$, in the $z$-direction. 
As we are interested in quantities perturbative in the
field strength, we can perform the analytic continuation to 
Minkowski space trivially~\cite{Tiburzi:2008ma}. 
The lattice simulations, moreover, 
are performed in Euclidean space for which the choice
in Eq.~\eqref{eq:AE} is natural. 
Throughout, we work with theories defined on a spatially periodic torus
of infinite time extent. 
The Euclidean space action has the form
\begin{equation}
S 
= 
\int_{-\infty}^\infty dx_4 
\int_0^L d\bm{x} 
\, \cL(x)
.\end{equation}
The length of each spatial direction is given by $L$. 
As the field is implemented by a time-dependent
gauge potential, there is no conflict between 
Eq.~\eqref{eq:AE} and the spatial periodicity of 
the underlying lattice action.%
\footnote{
In actual lattice simulations, the length of the time
direction is longer than the spatial directions, but obviously not infinite. 
To arrive at a periodic action in this case, the field strength must be 
quantized in the form 
$q \cE = 2 \pi n / \beta L$,
where $\b$ is the temporal extent of the lattice, 
and $n$ is an integer~\cite{'tHooft:1979uj,vanBaal:1982ag,Smit:1986fn,Rubinstein:1995hc}. 
We assume this quantization condition has been met, 
but that
$\b \gtrsim 2 L$ 
so that effects from the finite temporal extent of the lattice are comparatively quite small.  
}

The low-energy dynamics of pions is described by 
chiral perturbation theory~\cite{Gasser:1983yg}, 
in which pions are the Goldstone modes emerging
from spontaneous chiral symmetry breaking.%
\footnote{
At finite volume, we must also specify that the
volume is large enough to prevent the pion 
zero modes from conspiring to restore chiral symmetry. 
With this assumption, 
we work in the so-called $p$-regime~\cite{Gasser:1987ah,Leutwyler:1987ak,Gasser:1987zq}.
}
These modes are non-linearly realized in $\Sigma$, 
where
$\Sigma = \exp ( 2 i \phi / f )$,
and $\phi$ is an $SU(2)$ matrix of pion fields. 
The dimensionful parameter $f$ can be identified
with the chiral limit value for the pion decay constant, 
and $f = 130 \, \texttt{MeV}$  in our normalization.
The chiral Lagrangian appears as
\begin{equation} \label{eq:L}
\cL 
= 
\frac{f^2}{8}  < D_\mu \Sigma^\dagger D_\mu \Sigma > 
- 
\frac{\lambda}{2} < m_Q ( \Sigma^\dagger + \Sigma) >
,\end{equation}
where we have used a bracketed notation to denote
flavor traces: $< A > = \tr (A)$. The quark mass matrix
is $m_Q$, where $m_Q = \diag (m_u, m_d)$, 
and the action of the covariant derivative is specified by
\begin{equation}
D_\mu \Sigma = \partial_\mu  \Sigma + i e A_\mu [ \cQ, \Sigma ]
,\end{equation} 
where $\cQ$ is the quark electric charge matrix, $\cQ = \diag ( 2/3, - 1/3)$. 
In writing this Lagrangian, we have included only the lowest-order
terms. These terms scale as $\cO(\e^2)$, where 
$\e$ is a small dimensionless number. 
We assume the power counting:
$k^2 / \L_\chi^2 \sim m_\pi^2 / \L_\chi^2 \sim \e^2$. 
Here $\L_\chi$ is the chiral symmetry breaking scale, 
$\L_\chi = 2 \sqrt{2} \pi f$, $k$ is a typical 
momentum, and $m_\pi$ is the pion mass. 
The leading-order pion mass can be determined by 
expanding the Lagrangian in Eq.~\eqref{eq:L} 
to quadratic order in the meson fields.
Using the strong isospin limit, $m_u = m_d = m$, 
we have $f^2 m_\pi^2  = 4 \lambda m$.

%
\begin{figure}
\epsfig{file=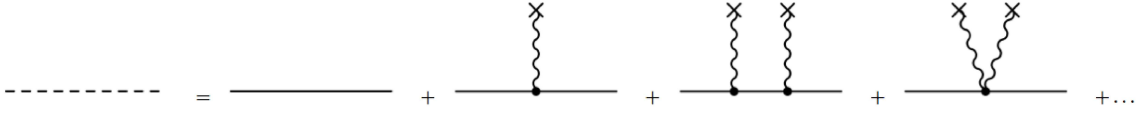,angle=270,width=15cm}
\caption{\label{f:prop} 
        Pion propagator in strong external fields. Depicted are the Born couplings to the charged pion 
        which have been summed to arrive at Eq.~\eqref{eq:pionprop}.}%
        \end{figure}
%

At leading order, 
the neutral pion propagator maintains a Klein-Gordon 
form due to the absence of charge couplings to the external field. 
For the charged pions, there are additional interactions.
Although we shall be interested in small fields, i.e.~$e F_{\mu \nu} / m_\pi^2 \sim \e^2$, 
it is actually easiest first to treat this ratio as $\cO(\e^0)$, 
and then expand quantities to the desired order in the external field strength. 
Summing the Born couplings to all orders, see Figure~\ref{f:prop}, 
produces the propagator of  the positively charged pion
\begin{equation} \label{eq:pionprop}
D(x',x)
=
\frac{1}{2 L^3} \int_0^\infty ds \sum_{\bm{n}} 
e^{ i \bm{k} \cdot (\bm{x}' - \bm{x})}
\Bigg\langle
x'_4 - \frac{k_3}{e \cE} , s 
\, \Bigg| \, 
x_4 - \frac{k_3}{e \cE} , 0 
\Bigg\rangle
e^{ - \frac{1}{2} s  E_{ \bm{k}_\perp}^2}
,\end{equation}
where $E_{\bm{k}_\perp}^2 = k_1^2 + k_2^2 + m_\pi^2$, 
and the periodic momentum modes are given in terms of triplets of integers $\bm{n}$, 
by the relation $\bm{k} = 2 \pi \bm{n} / L$.
This propagator depends on the quantum mechanical 
harmonic oscillator propagator evolved in proper-time $s$, 
\begin{equation}
\langle X' , s \, | \, X, 0 \rangle
=
\sqrt{\frac{e \cE}{2 \pi \sinh e \cE s}}
\exp 
\left\{
- \frac{e \cE}{2 \sinh e \cE s}
\left[
(X'^2 + X^2 ) \cosh e \cE s
- 2 X' X
\right]
\right\}
.\end{equation}

\begin{figure}
\epsfig{file=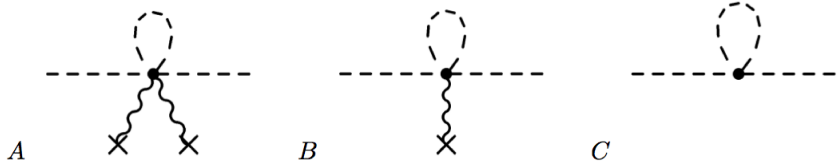,angle=270,width=11cm} 
        \caption{Diagrams for the pion two-point function in a non-perturbative external field.
        The dashed lines represent pion propagation in the background field. 
        The wiggly lines terminating in crosses represent 
        couplings of the vertices to the background field.}%
	\label{f:pion}
\end{figure}

To compute finite volume corrections to pion two-point
functions in background fields, we use the charged
pion propagator above in loop diagrams that arise 
in the effective theory, see Figure~\ref{f:pion}. 
The neutral pion contributions
to such loop diagrams only lead to field independent modifications.
These modifications are part of the
finite volume corrections to the pion mass.
Computation of charged pion loop diagrams with Eq.~\eqref{eq:pionprop}
leads to ultraviolet divergences, which we regulate using a simple subtraction
\begin{equation}
\sumint \equiv 
\frac{1}{L^3} \sum_{\bm{n}} - \int \frac{d \bm{k}}{(2 \pi)^3}
,\end{equation} 
that produces only the infrared corrections we seek.

%
%

It is instructive to consider the simplest loop contribution explicitly.  
The four-pion vertex arising from the mass term in Eq.~\eqref{eq:L}
makes contributions to diagram $C$ depicted in the figure. 
This mass-insertion tadpole diagram depends on the self-contracted propagator. 
Denoting the space-time coordinate of the vertex as $y_\mu$, 
we have the ultraviolet regulated pion self-contraction
\begin{eqnarray}
D(y,y) 
&\overset{\text{UV reg.}}{\longrightarrow}&
\d_L [D(y,y)] =
\frac{1}{2} \int_0^\infty ds \sumint 
\Bigg\langle
y_4 - \frac{k_3}{e \cE} , s 
\, \Bigg| \, 
y_4 - \frac{k_3}{e \cE} , 0 
\Bigg\rangle
e^{ - \frac{1}{2} s  E_{ \bm{k}_\perp}^2}
\\
&=&
\ldots +
\frac{e^2 \cE^2}{6}
\int_0^\infty ds \sumint 
\sqrt{\frac{s}{\pi}}
\left[
- 3 y_4^2 + s ( -1 + 6 k_3^2 y_4^2) + s^2 k_3^2 
\right]
e^{ - s [ \bm{k}^2 + m_\pi^2]}
+ \ldots
,\notag \\
\label{eq:masstadpole}
\end{eqnarray}
where in the second line we have series expanded in powers of the field strength $\cE$, 
keeping only the second-order term. The zeroth-order term in $\cE$ combines with the
corrections from neutral pion loops to give the finite volume pion mass shift. 
We assume that the background field strengths employed on the lattice are sufficiently 
small so that the fourth order and higher-order terms can be neglected. 
The sums in Eq.~\eqref{eq:masstadpole} are standard in the study of finite volume
modifications to meson properties. 
Using the catalog of finite volume functions of~\cite{Detmold:2006vu}, we can write the 
finite volume modification at $\cO(\cE^2)$ as 
\begin{equation}
\d_L [D(y,y)]
=
\frac{1}{4} e^2 \cE^2 
\left\{
y_4^2 
\left[
J_{5/2} (m_\pi) - I_{3/2}(m_\pi) 
\right]
+ 
\frac{5}{12} J_{7/2} (m_\pi)
- 
\frac{1}{2} I_{5/2} (m_\pi)
\right\}
\label{eq:masstadpoleanswer}
.\end{equation}
The last two terms give rise to a field-dependent 
shift in the effective mass squared. 
These are thus contributions to the finite volume
electric polarizability.

Notice the remaining dependence on the vertex time, $y_4$, in the first terms of Eq.~\eqref{eq:masstadpoleanswer}. 
In infinite volume, such dependence disappears
owing to an allowed shift of the $k_3$ integrand:
\begin{equation}
\int_{- \infty}^\infty  
d k_3  \,
e^{ - \alpha ( k_3 - e \cE y_4)^2}
= 
\int_{- \infty}^\infty  
dk_3 \,
e^{- \alpha k_3^2}
\notag
.\end{equation} 
At finite volume, discrete translational invariance
is not enough to remove the time-dependence.%
\footnote{
There is one special case: for a discrete torus
with field strengths quantized in the form $e \cE = 2 \pi n / L$
the time-dependence can be shifted away by reindexing
the sum on momentum modes. As these field
strengths are prohibitively large, however, we will not separately
address this case for which we would need an entirely 
different treatment. 
} 
This time dependence leads to an infrared renormalization
of the single-pion effective action, as we now demonstrate. 
Consider the full pion two-point function $G(x',x)$. 
Because we have discrete translational invariance in 
space, we can choose to locate the pion source at $\bm{x} = \bm{0}$. 
Furthermore, because the three momentum is a good 
quantum number, we can project onto a final-state pion at rest.
This leads us to
\begin{equation}
G(t,0) \equiv \int_0^L d \bm{x} \,  G(\bm{x}, t; \bm{0}, 0)
,\end{equation}
and similarly for the perturbative propagator, $D(t,0)$.
This notation allows us to focus on the time-dependence.
A perturbative contribution to $G(t,0)$
with time-dependence as in Eq.~\eqref{eq:masstadpoleanswer}
has the form
\begin{equation}
G(t,0) = D(t,0) - \cA \, \cE^2  \int dy_4 \, D(t,y_4) \, y_4^2 \, D(y_4,0)
.\end{equation}
The coefficient $\cA$ arises from the one-loop graphs;
and, as such, can be treated as a perturbation. 
Using a quantum mechanical notation, we have
\begin{eqnarray}
\langle t | G | 0 \rangle 
&=& 
\langle t | D | 0 \rangle
- 
\cA \,  \cE^2
\int dy_4 \, \langle t | D | y_4 \rangle \, y_4^2 \, \langle y_4 | D |0 \rangle
\notag \\
&=&
\langle t | \left[  D - \cA \,  \cE^2 D T^2 D  \right]  | 0 \rangle
=
\langle t | D \frac{1}{1 + \cA \,  \cE^2 T^2 D} | 0 \rangle + \ldots
,\end{eqnarray}
where $ T | t  \rangle = t | t \rangle$.
Applying the operator identity
\begin{equation}
\frac{1}{1+ A D} = D^{-1} \left[ D^{-1} + A \right]^{-1}
,\end{equation}
we see that the inverse propagator $G^{-1}$ 
has been additively renormalized
\begin{equation}
G^{-1} = D^{-1} + \cA \, \cE^2 T^2
.\end{equation}
For the case of charged and neutral pions
projected onto vanishing three-momentum, 
the inverse perturbative propagator has the 
form
\begin{equation} \label{eq:action}
D^{-1} 
= 
- \frac{\partial^2}{\partial T^2}
+ m_\pi^2
+ Q^2 \cE^2 T^2
,\end{equation}
where $Q$ is the pion charge. 
Hence the perturbation by $\cA$ corresponds
to an infrared renormalization of the field-squared
coupling:
$Q^2 \to Q^2 + \cA$. 
This is the coordinate space analog of the 
infrared renormalization of the zero-frequency 
Compton scattering tensor, originally found by~\cite{Detmold:2006vu}.
Gauge invariance is maintained, however, because
Ward identities no longer protect zero-frequency 
scattering~\cite{Hu:2007eb}. 
A gauge invariant operator that gives rise to a shift
in the charge-squared coupling of the neutral pion, for example, is
\begin{equation}
\cL 
= 
\frac{1}{2 L^2} \cA(L)  \,
\pi^0 \pi^0 
 \, \bm{W}^{(-)} \cdot \bm{W}^{(-)}
.\end{equation}
The $W_i^{(-)}$ is defined in terms of Wilson lines, 
$W_i^{(-)} =  \frac{1}{2 i} \left( W_i - W_i^\dagger \right)$,
where $W_i$ is given by
\begin{equation}
W_i = \exp \left( i \int_0^L dx_i \, A_i(x) \right)
.\end{equation}
The Einstein summation convention has been suspended 
in this equation. 
Because the space is compact, the Wilson line $W_i$ is actually
a gauge invariant loop. 
The coefficient of this operator, $\cA(L)$, runs to
zero exponentially fast as the volume is taken to infinity. 
Expanding the operator out to second order in the 
background field, we have
\begin{equation}
\cL = \frac{1}{2}
\cA(L) \,  \pi^0 \pi^0 \, \cE^2 t^2 + \cO(\cE^4)
,\end{equation}
which has precisely the form needed above
to describe renormalization of the charged-squared coupling.

The remaining contributions to the finite volume two-point function
arise from expanding out the kinetic term in Eq.~\eqref{eq:L}
to fourth order in the pion fields. The two gauge covariant
derivatives generate all three diagrams shown in Figure~\ref{f:pion}. 
The gauge part of the derivative leads to an explicit power of the 
external field at the vertex. Accordingly we can evaluate each diagram 
by using the propagator in Eq.~\eqref{eq:pionprop} in the appropriate contractions,
and expand the result to second order in the field strength. 
Unlike the infinite volume calculation, there are no particularly useful
cancellations to be aware of. 
Focusing on the sector with vanishing three-momentum, 
the resulting single-particle effective action to $\cO(\cE^2)$ has the form
\begin{equation} \label{eq:SeffE}
S_{\text{eff}}
= 
\int_{-\infty}^\infty dt \,\,
\pi^\dagger (t)
\left[
- \frac{\partial^2}{\partial t^2}
+ m_\pi^2(L)
+ 4 \pi \, m_\pi \,  \alpha_E(L) \, \cE^2
+ Q^2(L) \, e^2  \cE^2 t^2
\right]
\pi (t)
.\end{equation}
Here we have used $\alpha_E$ to denote the electric polarizability
which gives a positive contribution to the effective mass due to 
our Euclidean space treatment.
Each of the renormalized couplings is a sum of the infinite volume
piece plus a finite volume correction,
\begin{eqnarray}
m_\pi^2(L) &=& m_\pi^2 + \d_L[m_\pi^2], \notag  \\
\alpha_E(L) &=& \alpha_E + \d_L[\a_E],   \notag  \\
Q^2(L) &=& Q^2 + \d_L [Q^2]
.\end{eqnarray}

With the exception of the pion mass, which we do not consider 
here, the finite volume corrections are isospin dependent. 
For the neutral pion, we find
\begin{eqnarray}
\d_L[Q^2_{\pi^0}] &=& - \frac{1}{2 f^2} m_\pi^4 I_{5/2}(m_\pi)
,\\
\d_L[\a^{\pi^0}_E] &=& \frac{5 \, \a_{f.s} }{24 f^2 m_\pi} m_\pi^2 \left[ 2 I_{5/2}(m_\pi) +  J_{7/2}(m_\pi) \right]
,\end{eqnarray}
while for the charged pions, we have
\begin{eqnarray}
\d_L[Q^2_{\pi^\pm}] &=&- \frac{2}{3 f^2} \left[ 2 I_{1/2} (m_\pi) + m_\pi^2 I_{3/2}(m_\pi) \right]
,\\
\d_L[\a^{\pi^\pm}_E] &=& \frac{\a_{f.s.}}{3 f^2 m_\pi}  m_\pi^2 I_{5/2}(m_\pi)
.\end{eqnarray}
Here $\alpha_{f.s.} = e^2 / 4 \pi$ is the fine-structure constant.
For both charged and neutral pions, the values of $\d_L[Q^2]$
found here agree with those deduced from the zero-frequency Compton
scattering tensor~\cite{Hu:2007eb}. 
Focusing on the charged pion electric polarizability, 
we recall that delicate cancelations lead to only 
neutral pion loop contributions in infinite volume.
These contributions do not depend on the 
background field; thus, there are no one-loop effects
contributing to the polarizability. 
In finite volume, however, such cancelations are spoiled
due to the lack of SO(4) invariance. 
Consequently the value of $\d_L[\alpha^{\pi^\pm}_E]$ is non-vanishing.

%
%
%
%
%
\begin{figure}
\epsfig{file=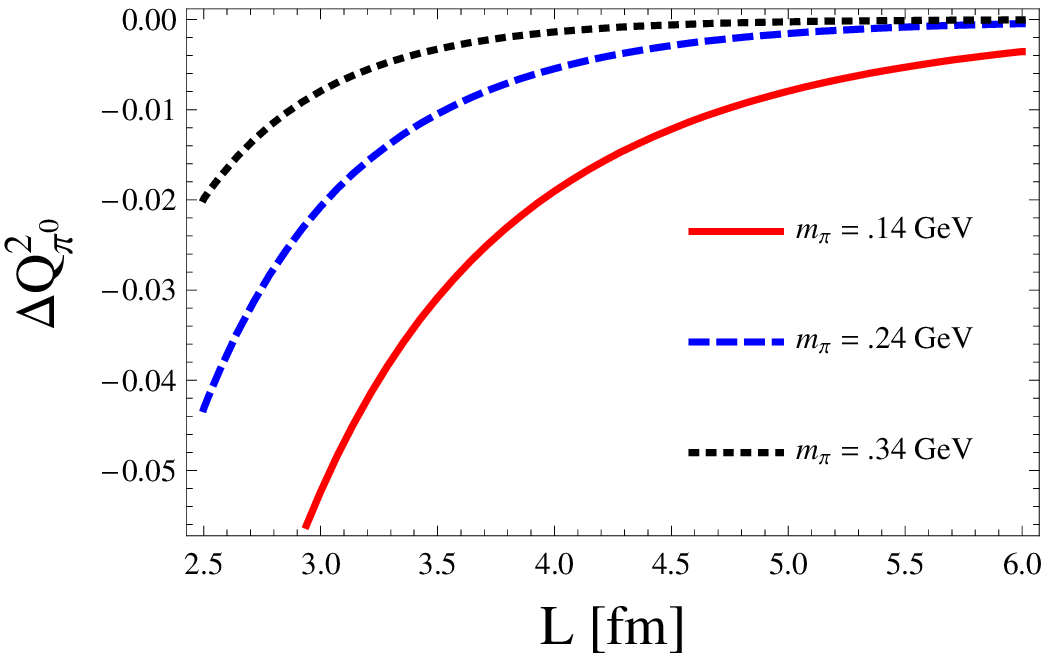,angle=0,width=7.8cm}
$\quad$
\epsfig{file=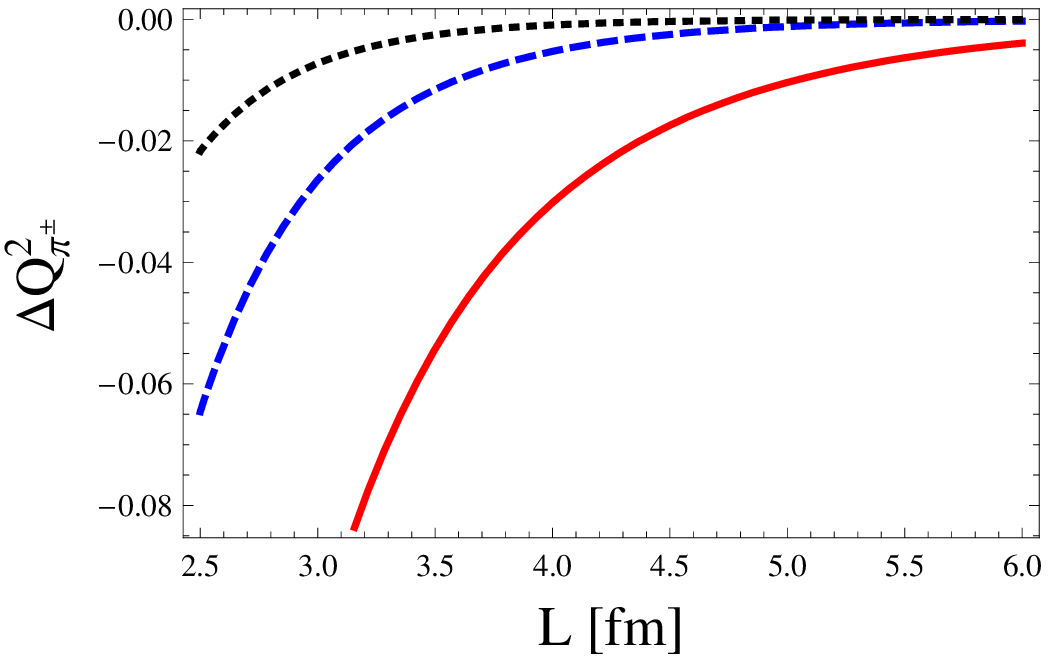,angle=0,width=7.8cm} 
        \caption{
                Relative change in charge-squared couplings. 
                Plotted versus $L$ for a few values of the pion mass 
                are the finite volume effects $\D Q^2$ for both neutral
                and charged pions. 
                }%
	\label{f:Qsquared}
\end{figure}
%
%
%
%
%

\begin{figure}
\epsfig{file=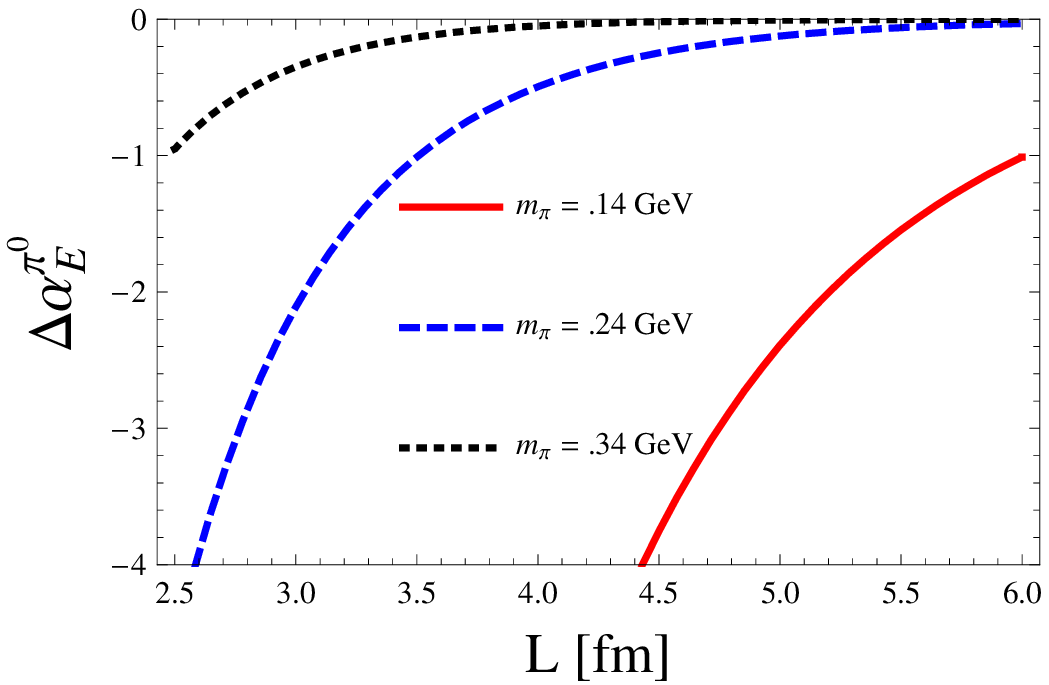,angle=0,width=7.8cm}
$\quad$
\epsfig{file=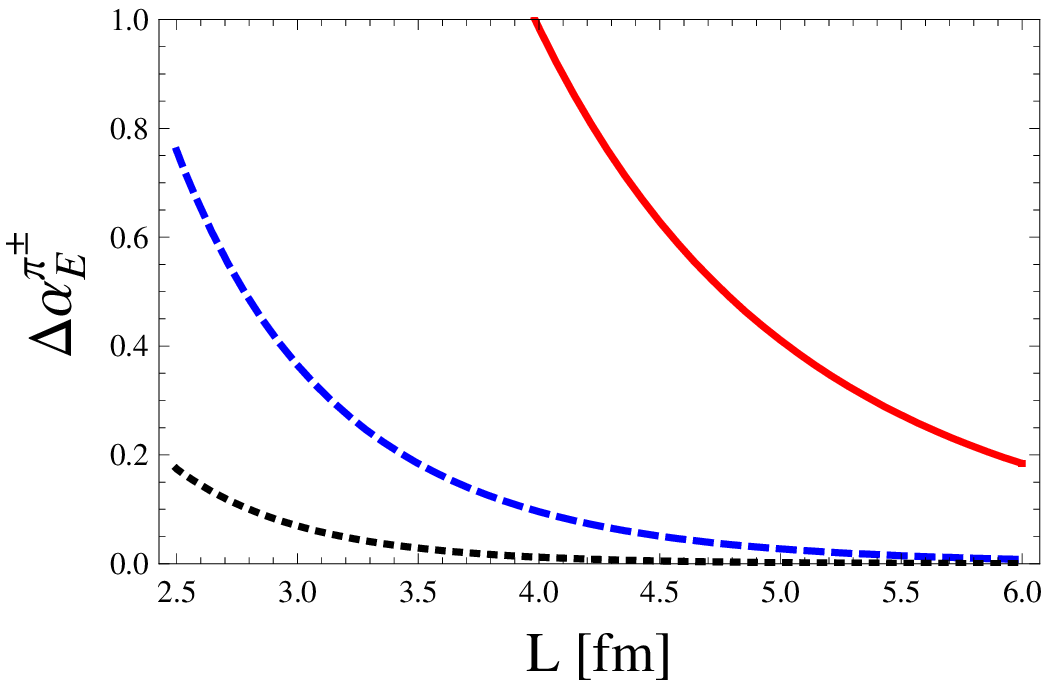,angle=0,width=7.8cm} 
        \caption{
                Relative change in electric polarizabilities. 
                Plotted versus $L$ for a few values of the pion mass 
                are the finite volume effects $\D \alpha_E$ for both neutral
                and charged pions. 
                }%
	\label{f:alphaE}
\end{figure}

To investigate the size of the finite volume corrections,
we consider the relative change in each of the quantities. 
For the charged pion, the relative change in charge squared
is just the finite volume effect,
$\D Q^2_{\pi^\pm} = \d_L[Q^2_{\pi^\pm}]$,
because in infinite volume $Q^2  =1$. 
For the neutral pion, we must define 
the relative change, since the infinite volume
charge-squared coupling is zero. 
We choose the simplest possible definition,
$\D Q^2_{\pi^0} = \d_L[Q^2_{\pi^0}]$. 
The charge-squared volume effects are
shown in Figure~\ref{f:Qsquared}. 
To consider the finite volume effects for the electric 
polarizabilities, we must recall the one-loop infinite volume
results~\cite{Bijnens:1987dc,Donoghue:1988ee,Holstein:1990qy}:
\begin{equation}
\alpha_E^{\pi^\pm} 
= 8 \, \a_{f.s} \frac{\a_9 + \a_{10}}{f^2 m_\pi} 
, 
\quad
\text{and}
\quad
\alpha_E^{\pi^0}
=
- \frac{\a_{f.s}}{3 (4 \pi f)^2 m_\pi}
.\end{equation}
Because both of these are non-vanishing, we
can define the relative change due to volume
effects in the ordinary manner,
$\D \alpha_E = \frac{\alpha_E(L) - \alpha_E}{\alpha_E}$.
To determine the low-energy constants required
for the charged pion polarizability, we use~\cite{Donoghue:1992dd}
\begin{equation}
\a_9 + \a_{10} = \frac{1}{32 \pi^2} f_A / f_V
,\end{equation}
where the form factors are given by $f_A = 0.0115$~\cite{Amsler:2008zz}, and $f_V = 0.0262$~\cite{Pocanic:2006jt}.
The latter value is consistent with the determination of 
$f_V$ assuming conservation of vector current, 
$f_V = 0.0259$. 
Notice the combination $\a_9 + \a_{10}$ is renormalization 
scale independent. 
The finite volume effect on the electric polarizabilities
is show in Figure~\ref{f:alphaE}, which depicts non-negligible corrections.
The relative change in neutral pion polarizability is more
sensitive to volume effects than the charged pion. One 
reason for this difference is the scaling by the infinite volume
result. The infinite volume polarizability is a factor of five times 
smaller for the neutral pion compared to the charged pion, 
while the finite volume modifications are roughly the same size.

%
%
%
%
%
%
%

The neutral pion polarizability is already challenging 
to calculate on the lattice due to disconnected quark
contractions (which have so far been neglected in simulations). 
In infinite volume, the one-loop expression for the neutral
pion polarizability stems entirely from such disconnected 
quark contractions~\cite{Hu:2007ts}. 
The electric polarizability of the charged pion can 
be determined from the lattice. Reports of a preliminary
calculation have been given in~\cite{Detmold:2008xk}.
At this stage, contributions from background field couplings
to the sea quarks have been omitted due to computational
restrictions.  
To address the omission of disconnected quark contractions
for neutral and charged pions, 
we turn to a partially quenched theory to differentiate
between valence and sea quarks~\cite{Sharpe:2000bc,Sharpe:2001fh}.
We perform the partially quenched finite volume
analysis with degenerate valence and sea quarks
using a partially quenched charge matrix that 
distinguishes between valence and sea quark charges~\cite{Tiburzi:2004mv,Detmold:2005pt}. 
The latter are set to zero. 
Additionally for the neutral pion, we perform the computation 
for only the connected part of the two-point function. 
The results are as follows:
the charge-squared couplings are renormalized by
\begin{eqnarray}
\d_L \left[  Q^2_{\pi^0} \right]_{\text{QEM}}^\text{connected}
=
0, \quad \text{and} \quad  
\d_L \left[  Q^2_{\pi^\pm} \right]_{\text{QEM}} 
=
\d_L \left[  Q^2_{\pi^\pm} \right]
,\end{eqnarray}
where the subscript $\text{QEM}$ denotes that we have quenched the
electromagnetic charges, and the superscript $\text{connected}$ denotes
that only the connected part of the correlator has been retained. 
The shifts in polarizabilities are given by
\begin{eqnarray}
\d_L 
[  
\a_E^{\pi^0}
]_{\text{QEM}}^\text{connected}
=
\d_L 
[ 
\a_E^{\pi^\pm} 
]_{\text{QEM}} 
=
\frac{5 \,  \a_{f.s.}}{ 27 f^2 m_\pi} m_\pi^2 I_{5/2} (m_\pi)  
,\end{eqnarray}
and are thus of comparable size to those with unquenched charges. 
Finite volume results for pion electric polarizabilities
suggest that as the pion mass is brought down on fixed sized
lattices, it will become increasingly challenging to isolate
the infinite volume physics. 
A thorough study of the pion mass and volume dependence
of lattice data will be required to extract pion polarizabilities.

\section{Mesons in Finite Volume: Magnetic Case}
\label{s:Magnetic}

Now we investigate the finite volume modifications 
to pion two-point functions for the case of a magnetic 
field.
To be concrete, we implement  the background field
with the gauge potential
\begin{equation} \label{eq:AB}
A_\mu(x) = (- B x_2, 0, 0, 0)
,\end{equation}
which corresponds to a magnetic field, $B$, in the $z$-direction.
On a spatial torus, the implementation of Eq.~\eqref{eq:AB}
generally breaks the periodicity of the lattice. 
To maintain discrete translational invariance of the lattice action, 
the field strength must be quantized in the form%
~\cite{'tHooft:1979uj,vanBaal:1982ag,Smit:1986fn,Rubinstein:1995hc}
\begin{equation}
q B = \frac{2 \pi n}{L^2}
,\end{equation}
where $n$ is an integer. 
Provided this condition is met, 
we can expand hadronic field operators in periodic momentum modes,
and treat them in a constant background field.

The Schwinger proper-time method cannot be directly utilized for the magnetic case
because the $y$-component of momentum is missing in the resummed propagator. 
Pions propagating in a background magnetic
field at finite volume, however, can be handled using the free particle 
propagator in coordinate space.
We return to Eq.~\eqref{eq:L} and treat the external field as a small perturbation, 
$B / m_\pi^2 \sim \e^2$. 
Thus to leading order, the external field dependence can be neglected,
and we have the coordinate space propagator
\begin{equation} \label{eq:pertprop}
D(x',x) = \frac{1}{L^3} \sum_{\bm{n}} \int \frac{d k_4}{2 \pi} \frac{ e^{ i \bm{k} \cdot ( \bm{x}' - \bm{x} )} e^{i k_4( x'_4 - x_4)} }{\bm{k}^2 + k_4^2 + m_\pi^2}
\end{equation}
for both charged and neutral pions. 
The spatial momentum modes are again quantized in the form $\bm{k} = 2 \pi \bm{n} / L$. 
The remaining field-dependent Born terms in Eq.~\eqref{eq:L}
must then be treated as coordinate-space interactions in perturbation theory. 
There are two types of interactions: a two-pion, linear background field coupling;
and, a two-pion, quadratic background field coupling.

%
\begin{figure}
\epsfig{file=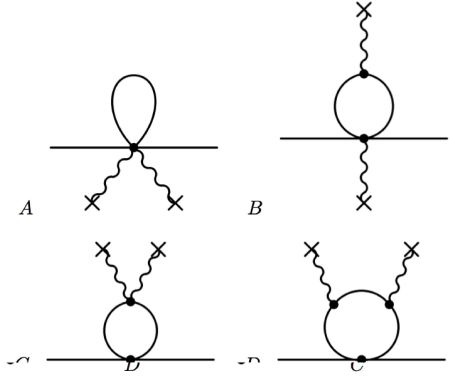,angle=270,width=5.75cm} 
        \caption{Diagrams for pion two-point functions in a perturbative external field.
        The straight lines represent free pion propagation.
        The wiggly lines terminating in crosses represent interactions with
        the background field.}%
	\label{f:pion2}
\end{figure}
%

To calculate one-loop corrections to pion two-point functions, 
we expand the Lagrangian in Eq.~\eqref{eq:L} to fourth order in the pion 
fields. We then generate all possible one-loop diagrams up to quadratic 
order in the external field $B$. Due to symmetry, diagrams linear in
the external field vanish.%
\footnote{
This is true when the action is projected onto vanishing 
$x$-component of momentum, $k_1=0$. 
For a general momentum, the infinite volume action contains a term
proportional to $2 i B x_2 k_1$, which acquires
infrared renormalization in finite volume.
This corresponds to screening of charged particle currents~\cite{Hu:2007eb}. 
Charge conjugation invariance forbids this effect for the neutral pion. 
}
The non-vanishing diagrams are shown 
in Figure~\ref{f:pion2}. It is particularly illustrative to consider a simple  
contribution explicitly. 
Consider the four-pion vertex with mass insertion derived from Eq.~\eqref{eq:L}. 
Using this vertex along with the quadratic background field coupling,
we arrive at a contribution to diagram $C$ in the figure. 
Let us denote the location of the four-pion vertex as $y_\mu$ and
the location of the quadratic field insertion as $z_\mu$. 
Then the correction to the propagator, 
$\d G(x',x)$, 
from this diagram is of the form
\begin{equation}
\d G(x',x) = - \int_{-\infty}^\infty  dy_4 \int_0^L d \bm{y} \, D(x',y) \D(y) D(y,x)
,\end{equation}
where, up to overall constants,
\begin{equation}
\D(y)  
=
- \int_{-\infty}^\infty dz_4 \int_0^L d\bm{z} \, D(y,z) \, z_2^2 B^2 \, D(z,y)
.\end{equation}
Inserting the perturbative propagator, Eq.~\eqref{eq:pertprop}, 
into the expression for $\D (y)$, 
yields three trivial integrations that can be performed: those over
$z_1$, $z_3$, and $z_4$. Clearly the $z_2$ integration is non-trivial given 
the coordinate dependence introduced by the gauge potential. 
After performing trivial integrations, we have
\begin{eqnarray}
\D (y) 
&=& 
- 
\frac{1}{L^4} 
B^2 
\int_0^L 
d z_2 
\int \frac{d k_4}{2 \pi} 
\sum_{\bm{n},m} 
\frac{
z_2^2 \, 
e^{ i k_2 ( y_2 - z_2)} 
e^{ i k'_2 ( z_2 - y_2)}
}
{
[ \bm{k}^2+ k_4^2 + m_\pi^2]
[k_1^2 + k_2^{'2} + k_3^2 + k_4^2 + m_\pi^2]
} \notag \\
&=& 
\frac{1}{L^4} B^2
\int_0^L 
d z_2 
\int \frac{d k_4}{2 \pi} 
\sum_{\bm{n},m} 
\frac{e^{- i (k_2 - k'_2) z_2}}{[k_1^2 + k_2^{'2} + k_3^2 + k_4^2 + m_\pi^2]}
\left(
\frac{\partial^2}{\partial k_2^2}
\frac{e^{i k_2 y_2}}{[\bm{k}^2 + k_4^2 + m_\pi^2]}
\right)
e^{ - i k'_2 y_2 }
,\notag \\
\end{eqnarray}
with $\bm{k} = 2 \pi \bm{n} / L$, and $k'_2 = 2 \pi m / L$. 
We then differentiate, perform the integrals over $z_2$ and $k'_2$ that have been rendered trivial, 
and drop terms that vanish by parity. 
Regulating the sum over momentum modes, we are left with
\begin{eqnarray}
\D (y) &=&
B^2 \int \frac{dk_4}{2 \pi} 
\sumint 
\left[
- \frac{y_2^2}{[ \bm{k}^2 + k_4^2 + m_\pi^2]^2}
- \frac{2}{[ \bm{k}^2 + k_4^2 + m_\pi^2]^3}
+ \frac{8 k_2^2}{[ \bm{k}^2 + k_4^2 + m_\pi^2]^4}
\right] \notag \\
&=&
B^2 \left[
- \frac{1}{4} y_2^2 \, I_{3/2}(m_\pi) 
- \frac{3}{8} I_{5/2}(m_\pi)
+ \frac{5}{12} J_{7/2}(m_\pi)
\right]
.\end{eqnarray}
The last two terms are finite volume corrections to the magnetic polarizability, 
while the first term is a finite volume correction to the charge-squared coupling.
The argument exposing this fact follows, \emph{mutatis mutandis}, from that for the electric case.

The remaining one-loop contributions to the pion two-point function 
can be evaluated similarly. The net result is that the pions are described
by an effective action of the form
\begin{equation} \label{eq:SeffB}
S_{\text{eff}} 
= 
\int_{-\infty}^\infty dt \int_0^L dy \, \, 
\pi^\dagger(y,t) 
\left[
- \frac{\partial^2}{\partial t^2}
- \frac{\partial^2}{\partial y^2}
+ m_\pi^2(L)
-  4 \pi \, m_\pi \,  \beta_M(L) \, B^2
+ Q^2(L) \, e^2  B^2 y^2
\right]
\pi(y,t)
,\end{equation}
up to $\cO(B^4)$ corrections. 
In writing this effective action, we have projected onto the sector with  $k_1 = k_3 = 0$.
The modification to the charge-squared term of the action 
is allowed by gauge invariance due to exactly the same operators
as in the electric case. The single-particle effective theory consequently demands
that the charge-squared couplings be identical to those determined above~\cite{Hu:2007eb}. 
The results of the explicit finite volume calculations indeed show 
that the $Q^2(L)$ in Eq.~\eqref{eq:SeffB} are identical to those in Eq.~\eqref{eq:SeffE}
for both $\pi^0$ and $\pi^\pm$. 
The same is additionally true for the case where
disconnected quark contractions are omitted.

%
%
%
\begin{figure}
\epsfig{file=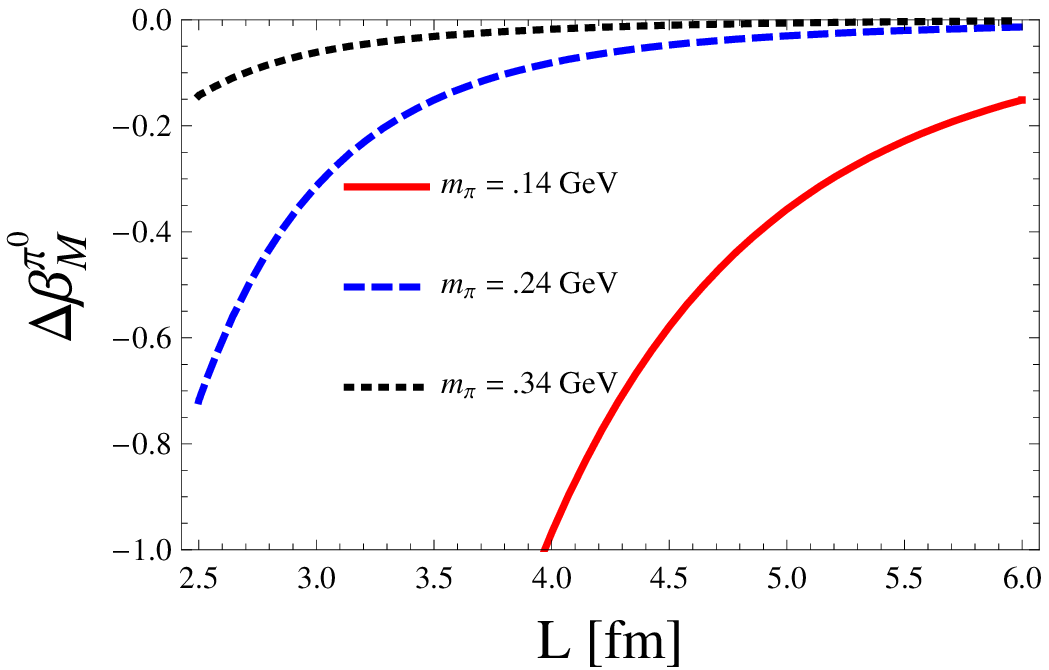,angle=0,width=7.8cm}
$\quad$
\epsfig{file=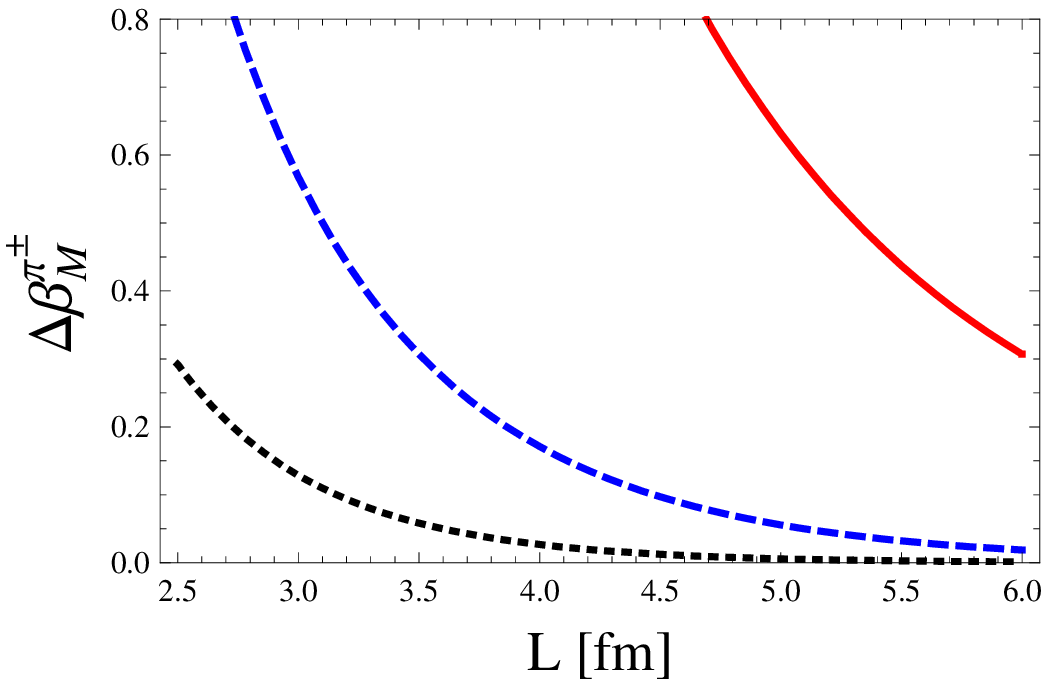,angle=0,width=7.8cm} 
        \caption{
                Relative change in magnetic polarizabilities. 
                Plotted versus $L$ for a few values of the pion mass 
                are the finite volume effects $\D \b_M$ for both neutral
                and charged pions. 
                }%
	\label{f:betaM}
\end{figure}
%
%
%
%

The term in the effective action with coefficient 
$\beta_M(L)$ 
denotes the magnetic polarizability contribution. 
This coefficient is a sum of the infinite volume polarizability,
and the finite volume modification
\begin{equation}
\b_M(L) = \b_M + \d_L[\b_M]
.\end{equation}
The former is given by 
$\b_M = - \a_E$ for charged and neutral pions, 
due to the helicity structure of the one-loop diagrams, 
while the latter are given by
\begin{eqnarray}
\d_L [ \b_M^{\pi^0} ] 
&=&
-
\frac{ \a_{f.s.}}{24 f^2 m_\pi}
m_\pi^2
\left[
2 I_{5/2}(m_\pi)
+ 
7 K_{9/2}(m_\pi)
\right],
\\
\d_L [ \b_M^{\pi^\pm} ] 
&=&
-
\frac{5 \, \a_{f.s}}{ 12 f^2 m_\pi}
m_\pi^4  I_{7/2} (m_\pi)
.\end{eqnarray}
The relative size of finite volume corrections to the magnetic 
polarizability,
$\D \b_M = \frac{\b_M(L) - \b_M}{\b_M}$, 
is investigated in Figure~\ref{f:betaM}.
The figure depicts non-negligible finite volume corrections. 
The corrections are of comparable size when
disconnected diagrams are neglected. 
Carrying out the partially quenched computation, we find
\begin{eqnarray}
\d_L 
[  
\b_M^{\pi^0}
]_{\text{QEM}}^\text{connected}
=
\d_L 
[ 
\b_M^{\pi^\pm} 
]_{\text{QEM}} 
=
-
\frac{25 \,  \a_{f.s.}}{ 108 f^2 m_\pi} m_\pi^4 I_{7/2} (m_\pi)  
.\end{eqnarray}

\section{Summary and Discussion}
\label{s:summy}

Above we have investigated finite size effects
on pion two-point correlation functions calculated in constant background fields. 
Treating the external field strength as weak, 
couplings to pions can be considered as perturbations in coordinate space.
One then uses coordinate-space perturbation theory on a torus to ascertain
the two-point correlation function in finite volume. 
In the case of electric fields, we utilized a time-dependent gauge 
potential which allowed us to sum the couplings to the external field 
to all orders. The resulting Schwinger proper-time Green's functions
were used to deduce volume corrections at one-loop order in chiral perturbation theory.
The effect of such corrections is to renormalize parameters of the single-particle effective
action in the infrared.
To second order in the electric field, we found the electric polarizabilities
and Born-level charge-squared couplings are renormalized. 
In the case of magnetic fields,
we treated the external field perturbatively from the outset,
and derived the single-particle effective action to second order
in the magnetic field. 
For this case, we found the infrared effect on the magnetic polarizabilities,
and an identical renormalization of charged-squared couplings
as in the electric case.
Numerical estimates of the finite size effects on polarizabilities show non-negligible 
dependence on the lattice size for pion masses $\lesssim 350\,  \texttt{MeV}$. 
Depending on the sign of such corrections relative to the infinite volume value, 
signals for polarizabilities will either be sizably enhanced, or sizably diminished. 
Based on one-loop chiral perturbation theory,  
the field-squared shifts in the effective mass of charged pions 
will be enhanced in both electric and magnetic fields. 
On the other hand, the shifts for neutral pions will 
be diminished in both electric and magnetic fields. 
Results were also presented for two-point functions calculated using
the lattice approximation of omitting the disconnected quark contractions. 
For the charged and neutral pions this approximation 
amounts to quenching the electric charges of the sea quarks; while for 
the neutral pion, the annihilation contraction is additionally neglected.

There are three clarifications that must be made about our results. 
Firstly and quite obviously, 
the gauge invariance of the single-particle effective actions is not manifest because
we have expanded results to second order in the external field. 
In infinite volume, the charge couplings are exactly fixed by Ward identities,
while in a compact space additional gauge invariant terms lead to renormalization
of charges and currents. 
Because we treat the time direction as infinite, charge is not renormalized. 
The possibility to renormalize spatial Born-level couplings in a gauge invariant manner
has been thoroughly detailed using a momentum-space analysis~\cite{Hu:2007eb}, 
and we reproduce the findings of that work using a complementary coordinate-space method. 
Finite volume corrections to polarizabilities show up as coefficients to field squared terms, $\bm{\cE}^2$ and $\bm{B}^2$.
These terms, moreover, can be further split into various gauge invariant terms. 
For example, at second order in the electric field strength one cannot distinguish
between operators of the form
\begin{equation}
\cO_1 =  \bm{\cE}\, {}^2
, \quad
\text{and}
\quad
\cO_2 =  \frac{\partial}{\partial x_4}  \bm{W}^{(-)} \cdot \bm{\cE}
,\end{equation}
because both terms have the form of polarizability operators at second order.
Another way to implement a constant electric field on a torus is to use the vector potential
\begin{equation} \label{eq:AEnot}
A_\mu(x) = (0,0,0,\cE x_3)
,\end{equation}
for which the operator $\cO_2$ vanishes. 
On a torus, however, the vector potentials, Eqs.~\eqref{eq:AE} and \eqref{eq:AEnot},
are not related by a gauge transformation. 
Consequently the finite volume shift of the polarizability 
(defined as the coefficient of the $\bm{\cE}^2$ term of the effective action)
will be different. 
A direction for future investigation is to classify all possible gauge invariant, 
single-particle operators allowed on a torus, 
and determine the running of their coefficients with the lattice volume. 
In this way, one could determine how to implement the external 
field with minimal volume corrections.

Secondly
the finite volume renormalization of charge-squared couplings must be addressed
to properly fit hadron correlation functions. Let us focus on the electric case.
For the charged pion, this is relatively straightforward. 
The two-point function arising from the effective action in Eq.~\eqref{eq:SeffE}
has precisely the form of Eq.~\eqref{eq:pionprop} projected onto zero three-momentum
with the mass replaced by the effective mass, namely
$m_\pi^2 \to m_\pi^2 + 4 \pi \alpha^{\pi\pm}_E(L) \, \cE^2$, 
and the charge-squared replaced by the renormalized charge-squared.
Because Eq.~\eqref{eq:pionprop} at zero momentum is an even function of the charge $e$,
the latter can be achieved simply by the trick
$e \to e \left( 1 + \frac{1}{2} \d_L[Q^2_{\pi^\pm}] \right)$.
While chiral perturbation theory estimates the latter at the few percent level, 
this expectation can be confirmed by treating the charge-squared as a free parameter
in fits to charged pion correlation functions.%
\footnote{ 
We stress that the charge (calculated from the time component of the current) has not been renormalized, 
as finite volume calculations indeed confirm~\cite{Hu:2007eb}. 
For the charged pion with non-zero momentum, 
the spatial current is screened in finite volume, 
and the effective charge arising from the spatial current is what appears in 
Eq~\eqref{eq:pionprop}. 
The square of this effective charge has been determined above
by projecting onto zero three-momentum.  
For the neutral pion, however, charge conjugation invariance forbids renormalization
of (odd powers of) the current operator(s). 
In the absence of current screening, however, 
a renormalization of the Thomson cross section gives rise to a 
charge-squared coupling. 
This is permitted because two current operators are involved. 
The finite volume neutral pion propagator at non-zero momentum
is then given by
\begin{equation}
D(x', x) = \frac{1}{2 L^3} \int_0^\infty ds \sum_{\bm{n}} e^{ i \bm{k} \cdot  ( \bm{x}' - \bm{x} )} 
\langle x'_4 , s | x_4, 0 \rangle e^{ - \frac{1}{2} s [\bm{k}^2 + m_\pi^2]}
\notag ,
\end{equation} 
where 
$\langle x'_4, s | x_4, 0 \rangle$ 
depends on the effective charge squared 
$\d_L[Q^2_{\pi^0}]$. 
}
The neutral pion, 
however, 
presents difficulty due to the sign of the renormalized charge-squared coupling.
The two-point function derived from Eq.~\eqref{eq:SeffE}
for the neutral pion appears as Eq.~\eqref{eq:pionprop} projected onto zero three-momentum, 
with 
$m_\pi^2 \to m_\pi^2 + 4 \pi \alpha^{\pi^0}_E(L) \, \cE^2$, 
and
$e \to i | Q_{\pi^0}(L)|$
necessitating analytic continuation.
The two-point function acquires an imaginary part
due to the instability in Eq.~\eqref{eq:SeffE} for negative charge squared.
This imaginary part is exponentially small
provided either the electric field strength is small,
$\cE / m_\pi^2 \ll 1$, 
or 
the pion Compton wavelength is small compared 
to the lattice size,
$m_\pi L \gg 1$. 
In practice, this should thus be a very small effect.
All neutral particles potentially have this difficulty. 
If the renormalized charge-squared is positive, 
we merely use Eq.~\eqref{eq:pionprop} with $Q(L)$,
and their correlation functions should look like the zero-momentum projection of charged particle correlation functions.
In particular, they will not be simple exponentials at long times. 
If, on the other hand, the renormalized charge-squared
is negative (as is the case for the neutral pion), 
our only recourse is to attempt to fit 
the correlation function using the real part of the 
analytically continued two-point function.
This procedure relies on the hope
that the imaginary part is numerically insignificant
to allow treatment in Euclidean space.
Notice for the connected part of the neutral pion 
two-point function, this issue does not arise
because $\d_L [ Q^2_{\pi^0} ]^{\text{connected}}_{\text{QEM}} = 0$.

Thirdly
the choice of coordinate origin is ordinarily arbitrary. 
In backgournd fields on a torus, however, this is no longer the case, 
as was pointed out for an electric field, see e.g.~\cite{Engelhardt:2007ub}.
Focusing on the electric case and the gauge potential in Eq.~\eqref{eq:AE}, 
we observe that two-point correlation functions
for both charged and neutral particles at finite volume will depend on the source time. 
In infinite volume this was already the case for charged particles via Eq.~\eqref{eq:pionprop}.
For neutral particles, charge-squared couplings are generated from infrared effects,
and a neutral hadron source will thus break discrete time-translational invariance.
Another way to shift the time is through the gauge potential in Eq.~\eqref{eq:AE}.
We are free to translate $x_4 \to x_4 + c$ in infinite volume
as it corresponds to a gauge transformation. 
In finite volume, this shift in time leads to an additional
coupling of the matter fields to a constant gauge potential
\begin{equation} \label{eq:AC}
A_\mu^{(c)}(x)  = ( 0 , 0 , - \cE c, 0 )
,\end{equation}
that cannot be gauged away on a torus. 
The effect is equivalent to a twisted boundary condition on the matter field, see e.g.~\cite{ZinnJustin:2002ru}.
Consequently a hadron with charge $Q$ acquires kinematic momentum,
$\bm{k} = - Q \cE c \hat{\bm{z}}$. 
One can extend the analysis formulated here to this flavor-twisted case
by working at non-zero spatial momentum.
Volume effects can be computed~\cite{Sachrajda:2004mi}, however, 
the expectation is that broken symmetries will lead to extra finite volume terms~\cite{Tiburzi:2006px,Jiang:2006gna,Jiang:2008ja}.
This complication notwithstanding, 
effects on the two-point function can be determined.
On general grounds, we expect all terms in the single particle effective action to be renormalized
including the current, or equivalently the momentum.
As was the case for periodic boundary conditions, neutral particle correlation
functions will no longer have a simple exponential falloff at long times.

Using the methods established here, 
many subsequent investigations are possible. 
Of particular interest is the calculation of finite volume
effects for baryon electromagnetic properties. 
These too can be addressed using the present framework.
In order to reliably extract these quantities from lattice QCD
simulations in background fields, 
one must ascertain whether volume corrections are as sizable 
as in the meson sector.

\begin{acknowledgments}
This work is supported in part by the 
U.S.~Department of Energy,
under
Grant No.~DE-FG02-93ER-40762.
\end{acknowledgments}

\appendix

\end{document}